\documentclass{iopart}

\begin{document}

\letter{Transverse frames for Petrov type I spacetimes: \\ a general algebraic procedure}
\author{Virginia Re\dag\ddag\S,\hspace{0.2cm} Marco Bruni\dag,\hspace{0.2cm} David R.\ Matravers\dag \hspace{0.2cm}and Frances T.\ White\dag}
\address{\dag\ Institute of Cosmology and Gravitation,\\
Portsmouth University, Portsmouth PO1 2EG, U.K.\\ \ddag\ Dipartimento di Fisica ``E.R. Caianiello",\\
Universit\`{a} di Salerno, via S.Allende 84081 Baronissi , Salerno, Italy\\ \S\ Istitituto Nazionale di Fisica
Nucleare, Sezione di Napoli}

\begin{abstract}
We develop an algebraic procedure to rotate a general Newman-Penrose tetrad in a Petrov type I spacetime into a
frame with Weyl scalars $\Psi_{1}$ and $\Psi_{3}$ equal to zero, assuming that initially all the Weyl scalars are
non vanishing. The new frame highlights the physical properties of the spacetime. In particular, in a Petrov Type
I spacetime, setting $\Psi_{1}$ and $\Psi_{3}$ to zero makes apparent the superposition of a Coulomb-type effect
$\Psi_{2}$ with transverse degrees of freedom $\Psi_{0}$ and $\Psi_{4}$.
\end{abstract}
 \pacs{Add}
 \hspace{2.4cm} {\footnotesize 29 November 2002} \\[0.7cm]

\section{Introduction}

The most promising sources of gravitational waves, for example the
merging of two black holes, are currently studied primarily using
numerical methods. Using the standard tools of classical
relativity to characterize the space-time in an invariant manner
we can complement this numerical work, contributing to the
physical interpretation of the output. There are many tools
developed for analyzing exact solutions which can be applied in
the context of numerical relativity.  Here we will work within the
Newman-Penrose formalism, on the problem of characterizing a
Petrov type I spacetime. Indeed one expects that realistic
gravitational wave sources will be described by a radiating
spacetime of type I.

In the Newman-Penrose formalism the ten independent components of
the Weyl tensor are expressed as  five complex scalars which are
the projection of the Weyl tensor on the null tetrad $\{
\mathbf{l},\mathbf{n},\mathbf{m},\mathbf{\bar{m}} \}$.  In vacuum,
all the physical information on the curvature is contained in
these scalars. Rotations of the tetrad, conveniently divided into
three classes, bring about transformations of these five Weyl
scalars (see for example \cite{Chand},\cite{exact}) that can be
used to produce simplifications.

The Petrov classification of the Weyl tensor depends on the number
of repeated roots of the equation $\Psi_{0}^{(New)}=0$, where
$\Psi_{0}^{(New)}$ is obtained by a class II rotation and is
expressed in terms of the initial scalars and the rotation
parameter \cite{Chand},\cite{exact}. When some of the roots
coincide the spacetime is said to be of special type. A pure plane
gravitational wave is Petrov type N (four  coincident roots) and
in this case the tetrad can be chosen to have $\Psi_{4}$ (or
$\Psi_{0}$) as the only non-zero Weyl scalar. The scalar
$\Psi_{2}$ represents a Coulomb-type tidal field. In a type D
spacetime (two distinct repeated roots) such as Schwarzschild or
Kerr  a tetrad can be found such that  $\Psi_{2}$ is the only
non-zero scalar. For Petrov type II spacetimes  (three distinct
roots, one of which is repeated)  a frame can be found where
$\Psi_{4}$ (or $\Psi_{0}$) and $\Psi_{2}$ are the only non-zero
scalars. In \cite{Pollney} a series of explicit algorithms for
putting the Weyl scalars into these standard forms, allowing for
all the possible initial non-zero components that one can have for
each special Petrov type, is presented.

The most general spacetime is Petrov type I, in which $\Psi_{0}^{(New)}=0$ has four distinct roots, and here the
physical meaning of the Weyl scalars can be highlighted if one chooses the tetrad so that the scalars appear in
one of the standard forms \cite{Pollney}:
\begin{equation}\label{sf1}
\{0,\Psi_{1},\Psi_{2},\Psi_{3},0\}
\end{equation}
\begin{equation}\label{sf2}
\{\Psi_{0},0,\Psi_{2},0,\Psi_{4}\}.
\end{equation}
The first case is the easier to achieve, at least in principle,
but the second case is physically more interesting in the light of
the physical interpretation of the Weyl scalars \cite{Szekeres,NP}
outlined above.  Since $\Psi_{1}$ and $\Psi_{3}$ represent
longitudinal ``gauge fields" they can conveniently be put to zero,
with $\Psi_{0}$ and $\Psi_{4}$ representing transverse degrees of
freedom, and $\Psi_{2}$ a Coulomb-type field.  In the following,
we call a tetrad in which the scalars take the form (2) a {\it
transverse frame}, and we shall denote the Weyl scalars in this
tetrad as $\Psi_0^{\top}$, $\Psi_2^{\top}$ and $\Psi_4^{\top}$
\cite{burko}.

If the spacetime is radiative the dominant Petrov type in the wave
zone will be N (Peeling-Off Theorem \cite{Sachs}), and one expects
$\Psi_0^{\top}$ and $\Psi_4^{\top}$ to represent gravitational
waves. If evaluated at a finite distance from the source,
$\Psi_0^{\top}$ and $\Psi_4^{\top}$ will  still be transverse, but
in general will include other effects. In relation to this, it is
worth pointing out that a Petrov type I spacetime does not
necessarily incorporate gravitational waves (for example there are
stationary metrics of type I), nor does the presence of
gravitational waves necessarily imply a non vanishing $\Psi_4$ or
$\Psi_0$, as it is clear from the existence of form (1) for any
type I spacetime that may include radiation.

Because of the physical interest of transverse frames we develop
here an algebraic procedure for arriving at a transverse tetrad
for a generic Petrov type I spacetime. We give expressions for the
final rotated scalars in terms of the initial scalars, which are
all assumed to be non vanishing. This is likely to happen if a
completely arbitrary choice of the initial tetrad frame is made.
The cases in which some of the initial scalars are zero in Petrov
type I are dealt with in \cite{Pollney}.

Section 2 is the main body of the letter, containing the details of the procedure.  Section 3 consists of
conclusions and discussion. The final expressions for the scalars that we obtain after performing the
transformations are shown in the Appendix.

\section{The Procedure}\label{sec2}
The problem for a given type I spacetime is how to arrive at a
transverse frame $\{\Psi_{0},0,\Psi_{2},0,\Psi_{4}\}$ from the
general case of all scalars non-zero. To do this we will use class
I and class II transformations, which  are \cite{Chand} :\\

\noindent \begin{tabular}{lll}

$ \mathbf{Class \hspace{2ex} I}$ & & $ \Psi_{0} \rightarrow  \Psi_{0}$ \\

& &  $ \Psi_{1} \rightarrow  \Psi_{1}+a^{*}\Psi_{0}$ \\

& &  $ \Psi_{2}\rightarrow \Psi_{2}+2a^{*}\Psi_{1}+(a^{*})^2\Psi_{0}$ \\

& &  $ \Psi_{3} \rightarrow \Psi_{3}+3a^{*}\Psi_{2}+3(a^{*})^2\Psi_{1}+(a^{*})^3\Psi_{0}$ \\

& &  $ \Psi_{4} \rightarrow \Psi_{4}+4a^{*}\Psi_{3}+6(a^{*})^2\Psi_{2}+4(a^{*})^3\Psi_{1}+(a^{*})^4\Psi_{0}$

\end{tabular} \\[0.5cm]

\noindent \begin{tabular}{lll}

$ \mathbf{Class \hspace{1ex} II}$ & & $ \Psi_{0} \rightarrow
\Psi_{0}+4b\Psi_{1}+6b^{2}\Psi_{2}+4b^{3}\Psi_{3}+b^{4}\Psi_{4} $ \\

& &  $ \Psi_{1} \rightarrow \Psi_{1}+3b\Psi_{2}+3{b}^{2}\Psi_{3}+b^{3}\Psi_{4} $ \\

& &  $ \Psi_{2}\rightarrow \Psi_{2}+2b\Psi_{3}+b^{2}\Psi_{4} $ \\

& &  $ \Psi_{3} \rightarrow \Psi_{3}+b\Psi_{4} $ \\

& &  $ \Psi_{4} \rightarrow  \Psi_{4} $

\end{tabular} \\

\noindent where $a$ and $b$ are complex parameters.

Starting from the general situation in which all the five scalars are different from zero, it is not easy to
obtain a transverse frame by directly carrying out the two types of rotations. To do this we would need to solve
the resulting system of equations for $a$ and $b$ :
\begin{eqnarray}
{\Psi_{1}}^{\top}&=&\Psi_{1}+b\Psi_{0}+3a(\Psi_{2}+2b\Psi_{1}+b^{2}\Psi_{0})  \nonumber \\
 & & {}+3a^{2}(\Psi_{3}+3b\Psi_{2}+3b^{2}\Psi_{1}+b^{3}\Psi_{0})  \nonumber \\
 & & {}+a^{3}(\Psi_{4}+4b\Psi_{3}+6b^{2}\Psi_{2}+4b^{3}\Psi_{1}+b^{4}\Psi_{0})=0 \\
{\Psi_{3}}^{\top}&=&\Psi_{3}+3b\Psi_{2}+3b^{2}\Psi_{1}+b^{3}\Psi_{0}  \nonumber \\
 & & {}+a(\Psi_{4}+4b\Psi_{3}+6b^{2}\Psi_{2}+4b^{3}\Psi_{1}+b^{4}\Psi_{0})=0
\end{eqnarray}

Here a class I transformation with parameter $b$ has been followed
by a class II transformation with parameter $a$. The difficulty is
that the degree of the above equations is too high for the problem
to be solved by a direct approach. In order to get around this, we
perform preliminary rotations on the original scalars to put
$\Psi_{4}=0$ and $\Psi_{0}=0$, arriving at the standard form (1).
This simplifies the system considerably because $\Psi_{0}$ is the
coefficient of $b^{4}$ and $\Psi_{4}$ is the coefficient of
$a^{4}$ in the transformation equation for the $\Psi$'s. So these
preliminary rotations ensure that the final equations to be solved
are only third order.

In total we perform four rotations:
\begin{center}$\Psi \stackrel{Class\hspace{1ex}I}{\rule[0.9mm]{2cm}{0.1mm}\hspace{-4mm}\rightarrow}
\Psi' \stackrel{Class\hspace{1ex}II}{\rule[0.9mm]{2cm}{0.1mm}\hspace{-4mm}\rightarrow} \Psi''
\stackrel{Class\hspace{1ex}I}{\rule[0.9mm]{2cm}{0.1mm}\hspace{-4mm}\rightarrow}\hspace{1ex}
\stackrel{Class\hspace{0.5em}II}{\rule[0.9mm]{2cm}{0.1mm}\hspace{-4mm}\rightarrow} \Psi^{\top} \nonumber$
\end{center}

\noindent where we require:
\begin{eqnarray}
\label{proced}
&\Psi&=\{\Psi_{0},\Psi_{1},\Psi_{2},\Psi_{3},\Psi_{4}\},\nonumber\\
&\Psi'&=\{\Psi_{0}',\Psi_{1}',\Psi_{2}',\Psi_{3}',0\},\\
&\Psi''&=\{0,\Psi_{1}'',\Psi_{2}'',\Psi_{3}'',0\},\nonumber \\
&{\Psi}^{\top}&=\{{\Psi}_{0}^{\top},0,{\Psi}_{2}^{\top},0,{\Psi}_{4}^{\top}\}.\nonumber
\end{eqnarray}

In order to get $\Psi_{4}'=0$, we apply a class I rotation, with parameter $b$, to the original scalars. The
expression for $\Psi_{4}'$ is a fourth order polynomial involving all the original scalars. This can be solved by
Maple. Substituting one of the solutions for $b$ into the rotation equations, we obtain expressions for the
rotated scalars in terms of the original ones. These turn out to be extremely long and must be manipulated before
they can be used in the further calculations.

The next step is to set $\Psi_{0}''$ equal to zero. We apply a class II rotation to the scalars $\Psi'$ (under
which $\Psi_{4}''=\Psi_{4}'$ remains zero) and require $\Psi_{0}''=0$.  This means solving a third order
polynomial equation in $a$. Choosing one of the solutions for $a$, we get the resulting non-zero scalars
($\Psi_{1}'',\Psi_{2}'',\Psi_{3}''$) in terms of the old ones.

Finally we rotate twice more, first class I (with parameter $g$) and then class II (with parameter $f$). Since in
our starting frame $\Psi_{0}''$ and $\Psi_{4}''$ are zero, the resulting equations for the new ${\Psi}_{1}^{\top}$
and ${\Psi}_{3}^{\top}$ are as follows:
\begin{eqnarray}
{\Psi_{1}}^{\top}&=&\Psi_{1}''+3f(\Psi_{2}''+2g\Psi_{1}'')+3f^{2}(\Psi_{3}''+3g\Psi_{2}''+3g^{2}\Psi_{1}'')
\nonumber\\ & &{}+ f^{3}(4g\Psi_{3}''+6g^{2}\Psi_{2}''+4g^{3}\Psi_{1}'')=0\\
{\Psi_{3}}^{\top}&=&\Psi_{3}''+3g\Psi_{2}''+3g^{2}\Psi_{1}''+f(4g\Psi_{3}''+6g^{2}\Psi_{1}''+4g^{3}\Psi_{1}'')=0
\end{eqnarray}
In this frame, the equations are only third order in both parameters and the system is straightforward to solve.
The solution is formally simple. For example the possible solutions include \cite{Pollney} :
\begin{eqnarray}
g=\sqrt{\frac{\Psi_{3}''}{\Psi_{1}''}},\hspace{1cm} f=1/g
\end{eqnarray}

However, the full expressions of the above quantities, in terms of the original scalars $\Psi$, are too long even
to be displayed. To make them more manageable, we implement the Maple command {\tt optimize}. This telescopes the
expressions by assigning to each sub-expression a name: $t_{1},t_{2},...$, etc and builds up a chain of
definitions relating these new objects. This greatly simplifies the implementation of these expressions in
computer language.

The final expressions for the scalars and all the definitions required for their compact expression are reported
in the Appendix. Here, for the sake of brevity, we report only the last line for each scalar:
\begin{eqnarray}
{\Psi}_{0}^{\top}&=&-2t_{140}/t_{143}+3/2t_{132}t_{140}(\Psi_{2}+t_{91}+t_{92}+t_{149}+2t_{143}t_{140})\nonumber
\\& & {} -1/2(12t_{79}+12t_{82}+4t_{86}+3t_{143}t_{160})/t_{155}\nonumber\\ & & {}
+1/16t_{167}(4t_{143}t_{87}+6t_{142}t_{160}+4t_{155}t_{140})/t_{97}\nonumber\\\\
{\Psi}_{2}^{\top}&=&\Psi_{2}+t_{80}+t_{82}+t_{136}+2t_{145}t_{142}\nonumber\\ & & {}
-(12t_{87}+12t_{89}+4t_{93}+3t_{145}t_{152})/t_{145}\nonumber
\\ & & {} +1/4t_{132}t_{142}(4t_{145}t_{94}+6t_{144}t_{152}+4t_{145}t_{144}t_{142})\nonumber \\\\
{\Psi}_{4}^{\top}&=&4t_{143}t_{87}+6t_{142}(\Psi_{2}+t_{91}+t_{92}+2t_{134}t_{87})+4t_{143}t_{142}t_{140}\nonumber\\
\end{eqnarray}

\noindent where the t's are temporary objects created by Maple, built from the original scalars.

\section{Conclusions}\label{sec4}

The transverse frame that we arrive at after our four rotations is
not unique, as there is a two-fold infinity of tetrads satisfying $\Psi_1$=$\Psi_3$=0.
This arbitrariness corresponds to the extra freedom that one has of performing  a further rotation
of class III, depending on two real parameters,  in order to completely fix the tetrad \cite{Pollney}.

The expressions for $\Psi_2$, $\Psi_0$ and $\Psi_4$ we obtain are quite complicated (see the Appendix). Partly they depend on the specific root of the polynomial one chooses at each step of the procedure (\ref{proced}). We have not investigated here wheater one particular path in this general scheme may bring about more manageable expressions for the scalars. In any  practical case  one will have to substitute the expression for the initial Weyl scalars in terms of the given metric and its derivative and these will be the expressions that one will want to simplify.

Nevertheless our  results are
completely general and can be used to put any Petrov type I
spacetime into the transverse frame in the case when  the
initial scalars are all non vanishing. This is useful whenever one
wants to investigate the physical properties of a Petrov Type I
spacetime using curvature variables.
A good example is the full non-linear numerical treatment of a
perturbed black hole. In this case the transverse frame is the
best way to highlight the perturbations affecting the curvature
field, which in the unperturbed case is represented by $\Psi_2$
only.

More generally,  it is of interest and potential value that our
{\tt optimized} expressions can in principle be applied to any
numerically computed spacetime in which one produces numerical
values for the initial scalars.

\section*{Acknowledgments}
We thank  Denis Pollney for useful suggestions, and Ray D'Inverno and Philippos Papadopoulos for discussions
during the initial stages of this work. VR thanks the ICG in Portsmouth for support during her visit. FW is
supported by a DTA grant from EPSRC. MB is partly supported by the EU programme `Improving the Human Research
Potential and the Socio--Economic Knowledge Base' (Research Training Network Contract HPRN--CT--2000--00137).

\section{Appendix}\label{Appendix}
Final expression for the rotated scalar ${\Psi_{0}}^{\top}$:

\begin{eqnarray}
t1 &=& \sqrt(6),\nonumber \\ t2 &=& \Psi_{1}^2,\nonumber \\ t3 &=&\Psi_{4}*\Psi_{2},\nonumber \\ t8 &=&
\Psi_{2}^2,\nonumber \\ t9 &=& t8*\Psi_{2},\nonumber \\ t11 &=& \sqrt(3), \nonumber \\t12 &=& \Psi_{4}^2,
\nonumber \\ t13 &=& \Psi_{0}^2,\nonumber \\ t16 &=& \Psi_{4}*t13,\nonumber \\ t19 &=& t8^2,\nonumber \\ t25 &=&
t2^2,\nonumber \\ t32 &=& \sqrt(\Psi_{4}*(-t12*t13*\Psi_{0}+18*t16*t8-81*\Psi_{0}*t19-54*t3*\Psi_{0}*t2+ \nonumber
\\ &+& 27*\Psi_{4}*t25+54*t2*t 9)), \nonumber \\ t36 &=&
(-27*t3*\Psi_{0}+27*\Psi_{4}*t2+27*t9+3*t11*t32)^{1/3},\nonumber \nonumber \\ t37 &=& t2*t36, \nonumber \\ t39 &=&
\Psi_{2}*\Psi_{0}, \nonumber \\ t40 &=& t39*t36, \nonumber
\\ t42 &=& t36^2,\nonumber \\ t48 &=& 1/t36,\nonumber
\\ t50
&=& \sqrt((6*t37-6*t40+t42*\Psi_{0}+3*t16+9*\Psi_{0}*t8)*t48),\nonumber
\\ t77
&=& \sqrt((72*t37*t50-72*t39*t36*t50-6*t50*t42*\Psi_{0}-\nonumber
\\ &+&18*t50*\Psi_{4}*t13-54*t50*\Psi_{0}*t
8-72*t2*\Psi_{1}*t1*t36+\nonumber \\ &+&108*\Psi_{1}*t1*t40)*t48/t50),\nonumber
\\ t78 &=& t1*t50+t77,\nonumber \\ t79 &=&
1/6*t78*\Psi_{2},\nonumber\\ t81 &=& 1/36*t78^2,\nonumber \\ t82 &=& t81*\Psi_{1},\nonumber \\ t86 &=&
1/6*t81*t78*\Psi_{0},\nonumber
\\ t87 &=& 3*t79+3*t82+t86,\nonumber \\ t88 &=& 1/6*t78*\Psi_{0},
\nonumber
\\ t89 &=& \Psi_{1}+t88,\nonumber \\ t90 &=& 1/6*t78*\Psi_{1},\nonumber \\ t91 &=& 2*t90,\nonumber \\
t92&=&t81*\Psi_{0},\nonumber \\ t93 &=&\Psi_{2}+t91+t92,\nonumber
\\ t94 &=& t89*t93,\nonumber \\ t97 &=& t87^2,\nonumber
\end{eqnarray}
\begin{eqnarray}
t100&=& t93^2,\nonumber\\ t101 &=& t100*t93,\nonumber \\ t103 &=& t89^2, \nonumber\\ t117 &=&
\sqrt(64*t103*t89*t87-36*t103*t100-108*t94*t87*\Psi_{0}+\nonumber
\\ &+&27*t13*t97+54*\Psi_{0}*t101 ),\nonumber \\ t122 &=&
(54*t94*t87-27*\Psi_{0}*t97-27*t101+3*t11*t117*t87)^(1/3),\nonumber
\\ t123 &=& 1/t122,\nonumber \\ t132 &=& 1/t87,\nonumber \\ t134 &=&
-2*t89*t123+(1/6*t122+3/2*t100*t123-1/2*\Psi_{2}-t90-1/2*t92)*t132,\nonumber
\\ t137 &=& t134^2,\nonumber \\ t140 &=& \Psi_{1}+t88+3*t134*t93+3*t137*t87,\nonumber \\ t142 &=&
t87/t140,\nonumber \\ t143 &=& \sqrt(t142),\nonumber \\ t149 &=& 2*t134*t87,\nonumber \\ t155 &=&
t143*t142,\nonumber \\ t160 &=& \Psi_{2}+t91+t92+t149,\nonumber \\ t167 &=& t140^2,\nonumber \\ t178 &=&
-2*t140/t143+3/2*t132*t140*(\Psi_{2}+t91+t92+t149+2*t143*t140)-\nonumber
\\&+&1/2*(12*t79+1
2*t82+4*t86+3*t143*t160)/t155\nonumber \\ &+&1/16*t167*(4*t143*t87+6*t142*t160+4*t155 *t140)/t97\nonumber
\end{eqnarray}\\

Optimized expression for the rotated scalar ${\Psi_{2}}^{\top}$ :
\begin{eqnarray}
t1 &=& \sqrt(6),\nonumber \\ t2 &=& \Psi_{1}^2,\nonumber \\ t3 &=& \Psi_{4}*\Psi_{2}, \nonumber \\ t8 &=&
\Psi_{2}^2, \nonumber \\ t9 &=& t8*\Psi_{2}, \nonumber \\ t11 &=& \sqrt(3), \nonumber \\ t12 &=& \Psi_{4}^2,
\nonumber \\ t13 &=& \Psi_{0}^2, \nonumber \\ t16 &=& \Psi_{4}*t13, \nonumber \\ t19 &=& t8^2, \nonumber \\ t25
&=& t2^2, \nonumber \\ t32 &=& \sqrt(\Psi_{4}*(-t12*t13*\Psi_{0}+18*t16*t8-81*\Psi_{0}*t19+\nonumber
\\ &-&54*t3*\Psi_{0}*t2+27*\Psi_{4}*t25+54*t2*t 9)), \nonumber \\ t36
&=& (-27*t3*\Psi_{0}+27*\Psi_{4}*t2+27*t9+3*t11*t32)^{1/3}, \nonumber \\ t37 &=& t2*t36, \nonumber \\ t39 &=&
\Psi_{2}*\Psi_{0}, \nonumber \\ t40 &=& t39*t36, \nonumber \\ t42 &=& t36^2, \nonumber \\ t48 &=& 1/t36, \nonumber
\\ t50 &=& \sqrt((6*t37-6*t40+t42*\Psi_{0}+3*t16+9*\Psi_{0}*t8)*t48), \nonumber \\ t77 &=&
\sqrt((72*t37*t50-72*t39*t36*t50-6*t50*t42*\Psi_{0}+\nonumber \\ &-&18*t50*\Psi_{4}*t13-54*t50*\Psi_{0}*t
8-72*t2*\Psi_{1}*t1*t36+\nonumber
\\&+&108*\Psi_{1}*t1*t40)*t48/t50), \nonumber
\\ t78 &=& t1*t50+t77, \nonumber \\ t79 &=& 1/6*t78*\Psi_{1}, \nonumber \\ t80 &=& 2*t79, \nonumber \\ t81 &=&
1/36*t78^2, \nonumber \\ t82 &=& t81*\Psi_{0}, \nonumber \\ t83 &=& 1/6*t78*\Psi_{0}, \nonumber \\ t84 &=&
\Psi_{1}+t83, \nonumber
\\ t85 &=& \Psi_{2}+t80+t82, \nonumber \\ t86 &=& t84*t85,
\nonumber
\\ t87 &=& 1/6*t78*\Psi_{2}, \nonumber \\ t89 &=& t81*\Psi_{1},
\nonumber
\\ t93 &=& 1/6*t81*t78*\Psi_{0}, \nonumber \\ t94 &=& 3*t87+3*t89+t93,
\nonumber \\ t97 &=& t94^2, \nonumber \\ t100 &=& t85^2, \nonumber
\\ t101 &=& t100*t85,\nonumber
\end{eqnarray}
\begin{eqnarray}
t103 &=& t84^2, \nonumber \\ t117 &=& \sqrt(64*t103*t84*t94-36*t103*t100-108*t86*t94*\Psi_{0}+\nonumber
\\ &+&27*t13*t97+54*\Psi_{0}*t101 ), \nonumber \\ t122 &=&
(54*t86*t94-27*\Psi_{0}*t97-27*t101+3*t11*t117*t94)^{1/3}, \nonumber \\ t123 &=& 1/t122, \nonumber \\ t132 &=&
1/t94, \nonumber
\\ t134 &=&
-2*t84*t123+(1/6*t122+3/2*t100*t123-1/2*\Psi_{2}-t79-1/2*t82)*t132, \nonumber \\ t136 &=& 2*t134*t94, \nonumber \\
t139 &=& t134^2, \nonumber \\ t142 &=& \Psi_{1}+t83+3*t134*t85+3*t139*t94, \nonumber
\\ t144 &=& t94/t142, \nonumber \\ t145 &=& \sqrt(t144), \nonumber \\ t152 &=&
\Psi_{2}+t80+t82+t136, \nonumber \\  t168 &=& \Psi_{2}+t80+t82+t136+2*t145*t142-\nonumber \\
&+&(12*t87+12*t89+4*t93+3*t145*t152)/t145+\nonumber \\&+& 1/4
*t132*t142*(4*t145*t94+6*t144*t152+4*t145*t144*t142)\nonumber
\end{eqnarray}\\

Optimized expression for the rotated scalar ${\Psi_{4}}^{\top}$ :
\begin{eqnarray}
t1 &=& \sqrt(6),\nonumber \\ t2 &=& \Psi_{1}^2,\nonumber \\ t3 &=&
\Psi_{4}*\Psi_{2}, \nonumber \\ t8 &=& \Psi_{2}^2, \nonumber \\ t9
&=& t8*\Psi_{2}, \nonumber \\ t11 &=& \sqrt(3), \nonumber \\ t12
&=& \Psi_{4}^2, \nonumber \\ t13 &=& \Psi_{0}^2, \nonumber \\ t16
&=& \Psi_{4}*t13, \nonumber \\ t19 &=& t8^2, \nonumber \\ t25 &=&
t2^2, \nonumber \\ t32 &=&
\sqrt(\Psi_{4}*(-t12*t13*\Psi_{0}+18*t16*t8-81*\Psi_{0}*t19+\nonumber
\\ &-&54*t3*\Psi_{0}*t2+27*\Psi_{4}*t25+54*t2*t 9)), \nonumber \\ t36
&=& (-27*t3*\Psi_{0}+27*\Psi_{4}*t2+27*t9+3*t11*t32)^{1/3},
\nonumber \\ t37 &=& t2*t36, \nonumber \\ t39 &=&
\Psi_{2}*\Psi_{0}, \nonumber \\ t40 &=& t39*t36, \nonumber \\ t42
&=& t36^2, \nonumber \\ t48 &=& 1/t36, \nonumber \\ t50 &=&
\sqrt((6*t37-6*t40+t42*\Psi_{0}+3*t16+9*\Psi_{0}*t8)*t48),
\nonumber \\ t77 &=&
\sqrt((72*t37*t50-72*t39*t36*t50-6*t50*t42*\Psi_{0}+\nonumber \\
&-&18*t50*\Psi_{4}*t13-54*t50*\Psi_{0}*t
8-72*t2*\Psi_{1}*t1*t36+\nonumber
\\&+&108*\Psi_{1}*t1*t40)*t48/t50), \nonumber
\\ t78 &=& t1*t50+t77, \nonumber \\ t79 &=& 1/6*t78*\Psi_{1}, \nonumber \\ t80 &=& 2*t79, \nonumber \\ t81 &=&
1/36*t78^2, \nonumber \\ t82 &=& t81*\Psi_{0}, \nonumber \\ t83
&=& 1/6*t78*\Psi_{0}, \nonumber \\ t84 &=& \Psi_{1}+t83, \nonumber
\\ t85 &=& \Psi_{2}+t80+t82, \nonumber \\ t86 &=& t84*t85,
\nonumber
\\ t87 &=& 1/6*t78*\Psi_{2}, \nonumber \\ t89 &=& t81*\Psi_{1},
\nonumber
\\ t93 &=& 1/6*t81*t78*\Psi_{0}, \nonumber \\ t94 &=& 3*t87+3*t89+t93,
\nonumber \\ t97 &=& t94^2, \nonumber \\ t100 &=& t85^2, \nonumber
\\ t101 &=& t100*t85,\nonumber
\end{eqnarray}
\begin{eqnarray}
t103 &=& t84^2, \nonumber \\ t117 &=&
\sqrt(64*t103*t84*t94-36*t103*t100-108*t86*t94*\Psi_{0}+\nonumber
\\ &+&27*t13*t97+54*\Psi_{0}*t101 ), \nonumber \\ t122 &=&
(54*t86*t94-27*\Psi_{0}*t97-27*t101+3*t11*t117*t94)^{1/3},
\nonumber \\ t123 &=& 1/t122, \nonumber \\ t132 &=& 1/t94,
\nonumber
\\ t134 &=&
-2*t84*t123+(1/6*t122+3/2*t100*t123-1/2*\Psi_{2}-t79-1/2*t82)*t132,
\nonumber \\ t136 &=& 2*t134*t94, \nonumber \\ t139 &=& t134^2,
\nonumber \\ t142 &=& \Psi_{1}+t83+3*t134*t85+3*t139*t94,
\nonumber
\\ t144 &=& t94/t142, \nonumber \\ t145 &=& \sqrt(t144), \nonumber \\ t152 &=&
\Psi_{2}+t80+t82+t136, \nonumber \\  t168 &=&
\Psi_{2}+t80+t82+t136+2*t145*t142-\nonumber \\
&+&(12*t87+12*t89+4*t93+3*t145*t152)/t145+\nonumber \\&+& 1/4
*t132*t142*(4*t145*t94+6*t144*t152+4*t145*t144*t142)\nonumber
\end{eqnarray}\\




\section*{References}
\begin{thebibliography}{99}

\bibitem{Chand} Chandrasekhar S {\it The Mathematical Theory of Black Holes}, Oxford
University Press

\bibitem{exact} Kramer D, Stephani H, Herlt E {\it Exact Solutions of Einstein's Field Equations}, Cambridge
University Press

\bibitem{Pollney}
D.Pollney, Skea J E F and d'Inverno R A, {\it Classifying geometries in General Relativity: I. Standard forms for
symmetric spinors}, Class. Quant.Grav.17 (2000).

\bibitem{Szekeres} Szekeres P, {\it{The gravitational compass}},
Jour.of Math.Phys. V6 (1965)

\bibitem{NP}
Newman E, Penrose R, {\it An approach to Gravitational radiation by a method of spin coefficients}, Jour.of
Math.Phys. V3 (1962)

\bibitem{burko}
Beetle C and Burko L, {\it A radiation scalar for Numerical Relatity}, gr-qc 0210019

\bibitem{Sachs} Sachs R, {\it{Gravitational waves in general relativity VI The outgoing radiation condition}},
Proc.Roy.Soc.(London)A264,309 (1961)


\end{thebibliography}
\end{document}